\documentclass[preprint]{revtex4}

\begin{document}

\title{Friedmann cosmology with decaying vacuum density}

\author{H. A. Borges and S. Carneiro}

\affiliation{Instituto de F\'{\i}sica, Universidade Federal da
Bahia, 40210-340, Salvador, BA, Brazil}

\begin{abstract}
Among the several proposals to solve the incompatibility between
the observed small value of the cosmological constant and the huge
value obtained by quantum field theories, we can find the idea of
a decaying vacuum energy density, leading from high values at
early times of universe evolution to the small value observed
nowadays. In this paper we consider a variation law for the vacuum
density recently proposed by Sch\"utzhold on the basis of quantum
field estimations in the curved, expanding background,
characterized by a vacuum density proportional to the Hubble
parameter. We show that, in the context of an isotropic and
homogeneous, spatially flat model, the corresponding solutions
retain the well established features of the standard cosmology,
and, in addition, are in accordance with the observed cosmological
parameters. Our scenario presents an initial phase dominated by
radiation, followed by a dust era long enough to permit structure
formation, and by an epoch dominated by the cosmological term,
which tends asymptotically to a de Sitter universe. Taking the
matter density equals to half of the vacuum energy density, as
suggested by observation, we obtain a universe age given by $Ht =
1.1$, and a decelerating parameter equals to $-1/2$.
\end{abstract}

\maketitle

\section{Introduction}

The huge difference between the small cosmological constant
inferred from observation and the vacuum energy density resulting
from quantum field theories has been, for a long time, a difficult
and fascinating problem for cosmologists and field theory
researchers \cite{Weinberg2,GRF}. Observations of the cosmic
background radiation indicates that we live in a spatially flat
universe \cite{Bernardis}, with total energy density equals, or
approximately equals, to the critical density. On the other hand,
gravitational measurements of matter density in the galaxies lead
to an average density of matter at the cosmological scale
approximately equals to one third of the critical density.
Therefore, we are led to the conclusion that two thirds of the
total energy are related to other, non-matter, component
\cite{omega}. Recent observations of type Ia supernovas at high
redshift \cite{supernovas} suggest that this dark energy component
exerts negative pressure, a natural candidate being then the
cosmological constant, which may be associated to the energy
density of the vacuum. Its value would then be $\Lambda \approx
10^{-52}$ m$^{-2}$ \cite{omega}.

On the other hand, when we calculate the energy density associated
with the vacuum quantum fluctuations, we obtain a divergent
result, which can be regulated by imposing an ultraviolet cutoff
of order of, say, the Planck mass (since at the Planck scale our
usual description of spacetime breaks down). In this case, we
obtain for the vacuum density the absurd figure $10^{122}$
m$^{-2}$ \cite{Weinberg2}. Even taking a smaller cutoff, as the
energies of the electroweak phase transition or the chiral
transition of QCD, we still have a very huge result compared to
the observed one. Furthermore, even dismissing the contribution of
the vacuum fluctuations, we still should deal with the vacuum
expectation value of any field undertaking phase transitions, as
the Higgs field, or the QCD condensate \cite{Weinberg2}.

A possible way out of this trouble is to consider a varying
cosmological term, which, as long as the universe expands, decays
from a huge value at initial times to the small value observed
nowadays \cite{Bertolami,Ozer,Freese}. Such an assumption can be
understood on the basis of a renormalization procedure, as
follows. The divergent vacuum energy density referred to above is
derived by using field theories in flat spacetime. On the other
hand, in the flat spacetime the left hand side of Einstein
equations is identically zero, which means that its right hand
side, that is, the total energy-momentum tensor, is also zero.
Therefore, in the Minkowski background the obtained divergent
result must be exactly canceled by introducing a bare cosmological
constant in the Einstein equations. Then, when we calculate the
vacuum energy density in a curved spacetime, we obtain again a
divergent result, but a finite, renormalized cosmological constant
follows by subtracting the Minkowskian divergent vacuum density.
In the case of an expanding background, this finite cosmological
term, being dependent on the curvature, should decay with the
expansion, being very large for initial times, and very small for
an almost flat universe like the ours.

To give a precise form for that reasoning, by making use of
quantum field theories in curved spacetimes, is a difficult
challenge. Nonetheless, an estimation has recently been made by
Sch\"utzhold \cite{Schutz}, who suggests that the main
contribution to the observed vacuum density arises from the QCD
trace anomaly, which, as he argues, dominates over the
contributions of the other sectors of the standard model of
particles interactions. This leads to a vacuum energy density
decaying as $\Lambda \approx m^3 H$, where $H$ is the Hubble
parameter, and $m \approx 150$ MeV is the energy scale of the
chiral phase transition of QCD. By using $H_0 \approx 70$
(km/s)/Mpc \cite{H}, it is easy to verify that this scaling law
leads to a present value in accordance with observation.

In this paper we investigate a cosmological scenario with such a
varying cosmological constant. We consider an isotropic and
homogeneous flat space, filled with matter and a cosmological term
proportional to $H$, obeying the equation of state of the vacuum.
We will show that the well established features of the standard
big-bang model are preserved, and that, in addition, the model is
in accordance with recent measurements of cosmological parameters
like the universe age, the deceleration parameter, and the
relative density between matter and vacuum energy. The universe
evolution has three distinct phases: a radiation dominated era,
with the same expansion rate as in standard cosmology; a phase
dominated by dust, long enough to permit structure formation; and
a later epoch dominated by the cosmological term, which tends
asymptotically to a de Sitter universe. As an additional feature
of the model, we show that the conservation of the total energy,
contained in the Einstein equations, leads to a process of matter
production, at the expenses of the decaying vacuum energy. This
process does not affect the primordial nucleosynthesis, and, for
late times like the ours, it is too small to be detected.

\section{Cosmological solutions with varying $\Lambda$}

Following the Sch\"utzhold suggestion, we will consider the
decaying vacuum energy density
\begin{equation} \label{Lambda}
\rho_{\Lambda} = \Lambda = \sigma H,
\end{equation}
where $\sigma$ is a positive constant of the order of $m^3$, and
we have made $8\pi G=c=\hbar=1$. Furthermore, we will take for the
vacuum the equation of state
\begin{equation} \label{pLambda}
p_{\Lambda} = - \rho_{\Lambda}.
\end{equation}
This is a natural choice: As the vacuum has the symmetry of the
background, its energy-momentum tensor has the form
$T^{\mu\nu}_{\Lambda} = \Lambda g^{\mu\nu}$, where $\Lambda$ is an
invariant function of the coordinates (in a homogeneous and
isotropic space, it is just a function of time). In comoving
coordinates, this corresponds to a perfect fluid with energy
density $\rho_{\Lambda}=\Lambda$, and pressure $p_{\Lambda} = -
\Lambda$.

As the matter component of the cosmic fluid, we will consider a
perfect fluid with equation of state
\begin{equation} \label{p}
p=(\omega - 1) \rho,
\end{equation}
where $p$ and $\rho$ are the pressure and energy density,
respectively. For dust matter ($p=0$) we have $\omega = 1$, while
for radiation ($p = \rho/3$) one has $\omega = 4/3$.

In the case of an isotropic and homogeneous, spatially flat
universe, the Einstein equations can be written as \cite{Weinberg}
\begin{eqnarray}
\label{Friedmann} \rho_T = 3H^2, \\ \label{continuidade}
\dot{\rho}_T + 3H (\rho_T + p_T) = 0,
\end{eqnarray}
where the dot means derivation with respect to the cosmological
time, and
\begin{eqnarray} \label{rhototal}
\rho_T &=& \rho + \rho_{\Lambda}, \\ \label{ptotal} p_T &=& p +
p_{\Lambda}
\end{eqnarray}
are the total energy density and total pressure, respectively.

Equation (\ref{Friedmann}), as well known, is the Friedmann
equation in the spatially flat case. In what concerns equation
(\ref{continuidade}), it is the continuity equation for the total
energy. Using (\ref{Lambda}), (\ref{pLambda}), (\ref{rhototal}),
and (\ref{ptotal}), it can be rewritten as
\begin{equation}
\dot{\rho} + 3H (\rho + p) = - \dot{\Lambda}.
\end{equation}
In the case of a constant $\Lambda$, we recover the continuity
equation for matter. The above equation shows us that, in order to
satisfy the energy conservation, a decaying vacuum term
necessarily leads to matter production \cite{Bertolami2}. The
variation in the number of particles is a known property of
non-stationary backgrounds. Its microscopic description is, in
general, a difficult task, for it involves quantum field
calculations in curved spacetimes. Here we will consider matter
production from a macroscopic perspective, obtaining its rate from
the Einstein equations.

The set (\ref{Lambda})-(\ref{ptotal}) leads us to the differential
equation
\begin{equation}
2\dot{H} + 3\omega H^2 - \sigma \omega H = 0,
\end{equation}
which determines the time evolution of the Hubble parameter. Apart
from an integration constant related to the choice of the origin
of time, its general solution is given by
\begin{equation} \label{geral}
t = \frac{2}{\sigma \omega} \ln \left| \frac{3H}{3H-\sigma}
\right|.
\end{equation}

From equations (\ref{rhototal}), (\ref{Friedmann}), and
(\ref{Lambda}), it is possible to verify that
\begin{equation} \label{rho}
\rho = (3H - \sigma) H.
\end{equation}
As the weak energy condition requires $\rho \geq 0$, and since, in
an expanding universe, we have $H = \dot{a}/a \geq 0$, it follows
that $3H - \sigma \geq 0$. Therefore, the solution (\ref{geral})
can simply be written as
\begin{equation} \label{t}
t = \frac{2}{\sigma \omega} \ln \left( \frac{H}{H-\sigma/3}
\right),
\end{equation}
leading to
\begin{equation} \label{H}
H = \frac{\sigma/3}{1-\exp(-\sigma \omega t/2)}.
\end{equation}

Integrating once more with respect to time, we obtain the scale
factor
\begin{equation} \label{a}
a = C \left[\exp\left(\sigma \omega t/2\right) -
1\right]^{\frac{2}{3\omega}},
\end{equation}
where $C$ is an integration constant.

Substituting (\ref{H}) into (\ref{Lambda}) and (\ref{rho}), we
obtain, respectively, the cosmological term and the matter density
as functions of time,
\begin{equation} \label{Lambdat}
\Lambda = \frac{\sigma^2/3}{1-\exp(-\sigma \omega t/2)}, \\
\end{equation}
\begin{equation}
\label{rhot} \rho = \frac{\sigma^2}{12}
\sinh^{-2}\left(\sigma\omega t/4\right).
\end{equation}
Therefore, the ratio between the vacuum and matter densities
scales as
\begin{equation}\label{Omega}
\Omega \equiv \frac{\Lambda}{\rho} = \exp(\sigma \omega t/2)-1.
\end{equation}
With the help of (\ref{t}) or (\ref{H}), it is possible to derive
from the above equation the relation
\begin{equation}\label{sigma}
\sigma = \frac{3H\Omega}{\Omega + 1}
\end{equation}
(which can also be found by using (\ref{Lambda}),
(\ref{Friedmann}), and (\ref{rhototal})). This equation allows one
to obtain a precise value for $\sigma$ from the observed current
values of $H$ and $\Omega$.

The vacuum and matter densities can also be expressed as functions
of the scale factor. With the help of (\ref{a}), we rewrite
equation (\ref{rhot}) in the form
\begin{equation}\label{rhoa}
\rho =
\frac{\sigma^2}{3}\left(\frac{C}{a}\right)^{3\omega/2}\left[1 +
\left(\frac{C}{a}\right)^{3\omega/2}\right],
\end{equation}
while for $\Lambda$ we have
\begin{equation}\label{Lambdaa}
\Lambda = \frac{\sigma^2}{3}\left[1 +
\left(\frac{C}{a}\right)^{3\omega/2}\right].
\end{equation}

From (\ref{a}) one can deduce the deceleration factor,
$q=\ddot{a}a/\dot{a}^2$. It is given by
\begin{equation}\label{q}
q = \frac{3\omega}{2} \exp(-\sigma \omega t/2) -1,
\end{equation}
which, by using (\ref{Omega}), can also be written as
\begin{equation}\label{qOmega}
q = \frac{3\omega}{2(\Omega+1)}-1.
\end{equation}

\section{The radiation era}

For the radiation epoch, we have $\omega = 4/3$. In this case,
equation (\ref{a}) is written as
\begin{equation}\label{aradiation}
a = C \left[\exp\left(2\sigma t/3\right) - 1\right]^{1/2}.
\end{equation}
In the limit of small time ($\sigma t\ll 1$), this expression
reduces to
\begin{equation}\label{asmall}
a \approx \sqrt{2C^2\sigma t/3}.
\end{equation}

On the other hand, expressions (\ref{rhoa})-(\ref{Lambdaa}) turn
out to be
\begin{equation}\label{rhoaradiation}
\rho = \frac{\sigma^2 C^4}{3a^4} + \frac{\sigma^2 C^2}{3a^2},
\end{equation}
\begin{equation}\label{Lambdaaradiation}
\Lambda = \frac{\sigma^2}{3} + \frac{\sigma^2 C^2}{3a^2}.
\end{equation}
In the limit $a \rightarrow 0$, they reduce to
\begin{equation}\label{rhosmall}
\rho = \frac{\sigma^2 C^4}{3a^4} = \frac{3}{4t^2},
\end{equation}
\begin{equation}\label{Lambdasmall}
\Lambda = \frac{\sigma^2 C^2}{3a^2} = \frac{\sigma}{2t}
\end{equation}
(where we have also used (\ref{asmall})).

From (\ref{asmall}) and (\ref{rhosmall}) one can see that the
scale factor and the matter density have the same time dependence
as in the standard model, and that the radiation density scales as
$a^{-4}$, as should be \cite{Weinberg}. Alternatively, comparing
(\ref{Lambdasmall}) to (\ref{Lambda}) we obtain $Ht=1/2$, and,
from (\ref{q}), with $t \rightarrow 0$ and $\omega=4/3$, we have
$q \approx 1$, again in accordance with the standard model.
Furthermore, comparing (\ref{rhosmall}) to (\ref{Lambdasmall}), we
see that, for $a \rightarrow 0$, the radiation density diverges
faster than the cosmological term, leading to $\Omega \approx 0$.
In other words, at early times the expansion is completely
dominated by radiation.

The first term of equation (\ref{rhoaradiation}) gives the usual
scaling of the radiation density. The second term is owing to the
process of matter production, resulting from the decay of the
vacuum energy density, as already discussed. Since, in the limit
of small time, the first term dominates, we expect that the matter
production does not interfere on processes taking place at early
times, as nucleosynthesis, for example. Let us show that this is
the case.

The rate of matter production can be defined in this context as
\begin{equation}\label{Tgamma}
T_{\gamma} = \frac{1}{\rho a^4} \frac{d}{dt} (\rho a^4)
\end{equation}
(in the case of a genuine cosmological constant, or in the absence
of any cosmological term, $\rho a^4$ is a constant, and this rate
is equal to zero). With the help of (\ref{rhoaradiation}),
(\ref{aradiation}), (\ref{sigma}), and (\ref{Omega}) (with $\omega
= 4/3$), it is possible to verify that
\begin{equation}
T_{\gamma} = 2\sigma/3 = \frac{2\Omega}{1+\Omega} H.
\end{equation}

We then obtain a constant rate. However, what is really important
in what concerns processes like nucleosynthesis is the ratio
between the rate of matter production and the expansion rate, that
is, the ratio
\begin{equation}
T_{\gamma}/H = \frac{2\Omega}{1+\Omega}.
\end{equation}
We have seen that, for early times, $\Omega \approx 0$, leading to
$T_{\gamma}/H \approx 0$. As the ratio between the nucleosynthesis
rate and the expansion rate is finite at the time of primordial
nucleosynthesis, we conclude that the matter production does not
affect this process, as expected.

To close this section let us comment that, although $\Lambda$
tends to infinity for small times, we do not have inflation in
this model (because the radiation density tends to infinity
faster). Nevertheless, let us recall that, in his derivation of
$\Lambda = \sigma H$ \cite{Schutz}, Sch\"utzhold has used the
approximation that the cosmological time scale, given by $H^{-1}$,
is very large compared to the time scale of the vacuum quantum
fluctuations. This approximation does not hold for very early
times, when inflation occur.

\section{Dust era and the limit of late times}

Let us now discuss the phase dominated by dust matter, that is,
the case $\omega = 1$. Now, the scale factor (\ref{a}) has the
form
\begin{equation} \label{adust}
a = C \left[\exp\left(\sigma t/2\right) - 1\right]^{2/3}
\end{equation}
(evidently, the integration constant is not the same as in
equations (\ref{aradiation})-(\ref{asmall})).

For small times (small compared to the present time), it can be
approximated by
\begin{equation}\label{adustsmall}
a = C(\sigma t/2)^{2/3},
\end{equation}
which has the same time dependence as in the standard flat model
with dust \cite{Weinberg}. Therefore, the radiation era is
followed by an epoch with decelerating expansion, as necessary in
order to allow structure formation. As we shall see, the varying
cosmological term starts dominating just at the present time,
which guarantees a large enough dust era.

From expressions (\ref{rhoa})-(\ref{Lambdaa}) (with $\omega=1$),
we obtain for the matter density
\begin{equation}\label{rhodust}
\rho = \frac{\sigma^2 C^3}{3a^3} + \frac{\sigma^2
C^{3/2}}{3a^{3/2}},
\end{equation}
while for $\Lambda$ one has
\begin{equation}\label{Lambdadust}
\Lambda = \frac{\sigma^2}{3} + \frac{\sigma^2 C^{3/2}}{3a^{3/2}}.
\end{equation}
The first term in (\ref{rhodust}) gives the usual scaling of dust
matter. The second term is related, as before, to the production
of matter, at the expenses of the vacuum decay.

In the limit of large times, that is, $\sigma t \gg 1$ and $a
\rightarrow \infty$, equations (\ref{adust}), (\ref{rhodust}), and
(\ref{Lambdadust}) lead to
\begin{eqnarray} \label{alarge}
a &=& C \exp\left(\sigma t/3\right),\\ \Lambda &=& \sigma^2/3, \\
\rho &\approx& 0.
\end{eqnarray}
That is, as long as the cosmological term tends asymptotically to
a genuine cosmological constant, our solution tends to a de Sitter
universe, with $H=\sqrt{\Lambda/3}=\sigma/3$. The same result can
be seen from equation (\ref{q}): In the limit $t \rightarrow
\infty$, it gives $q = -1$, which characterizes the de Sitter
solution \cite{Weinberg}.

Let us now obtain the universe age, i.e., the value of the
cosmological time at present. For any time, by using (\ref{sigma})
and (\ref{t}) (with $\omega=1$) it is easy to show that
\begin{equation}
t H = \frac{2(\Omega + 1)}{3\Omega} \ln(\Omega +1).
\end{equation}
The observations suggest that the present ratio between the vacuum
and matter densities is  $\Omega_0 \approx 2$ \cite{omega}. With
this value we have, from the above equation, $t_0H_0 \approx \ln 3
\approx 1.1$. This is inside the current limits for the universe
age, $0.8 \lesssim t_0H_0 \lesssim 1.3$, and in good accordance
with the best estimation $t_0H_0 \approx 1$ \cite{age}.

From equation (\ref{qOmega}), we can also obtain the present value
of the deceleration parameter. Taking $\omega = 1$ and $\Omega_0
\approx 2$, one has $q_0 \approx - 1/2$. This is the same result
obtained if we consider a genuine cosmological constant, and it is
consistent with supernova observations.

As commented above, the epoch dominated by dust, with decelerating
expansion, must be sufficiently large to allow the formation of
large structures. Let us verify that this is indeed the case. The
time $t_{\Lambda}$ for which the cosmological term starts
dominating is given by the condition $\Lambda = \rho$. Then,
equating (\ref{Lambdat}) to (\ref{rhot}) (with $\omega = 1$), and
using (\ref{sigma}), it is not difficult to derive
\begin{equation}
t_{\Lambda} = \frac{2\ln2}{3H_0}
\left(\frac{1+\Omega_0}{\Omega_0}\right).
\end{equation}
By using $\Omega_0 \approx 2$ and our previous result $t_0H_0
\approx \ln3$, we obtain $t_{\Lambda} \approx 0.6 \; t_0$.

On the other hand, we can also calculate the time $t_q$ for which
$q=0$, that is, when the expansion changes from the decelerating
phase to an accelerating one. Equating to zero the expression
(\ref{q}), with $\omega=1$, and using again (\ref{sigma}), we
obtain
\begin{equation}
t_q = \frac{2\ln(3/2)}{3H_0}
\left(\frac{1+\Omega_0}{\Omega_0}\right).
\end{equation}
With $t_0H_0 \approx \ln3$, one has $t_q \approx 0.4 \; t_0$.
Thus, the characteristic times $t_{\Lambda}$ and $t_q$ are close
to the present time, indicating that we are living just at the end
of the dust era.

Finally, let us investigate the production of matter at late
times. Analogously to the case of radiation (see eq.
(\ref{Tgamma})), the rate of matter production is now defined as
\begin{equation}
T=\frac{1}{\rho a^3} \frac{d}{dt} (\rho a^3)
\end{equation}
(when $T=0$, we have $\rho a^3=$ constant, as should be for
conserved dust matter). It can be calculated with the help of
equations (\ref{rhodust}), (\ref{adust}), (\ref{sigma}), and
(\ref{Omega}) (with $\omega = 1$), leading to
\begin{equation}
T = \sigma/2 = \frac{3\Omega}{2(1+\Omega)}H.
\end{equation}

In the limit of large times we have $\Omega \rightarrow \infty$,
and so
\begin{equation}
T_{\infty} = 3H/2.
\end{equation}
For the present time, on the other hand, we have $\Omega \approx
2$, and then
\begin{equation}
T_0 \approx H_0.
\end{equation}
These results (which, actually, are both equals to $\sigma/2$) are
smaller than the rate characteristic of the old stead-state
cosmology, given by $T = 3H$. They are beyond the current
possibilities of direct observation.

\section{Concluding remarks}

The proposal of a cosmological term scaling with $H$, which
follows from quantum field estimations of the vacuum energy
density in an expanding background \cite{Schutz}, besides to give
the small $\Lambda$ observed nowadays, leads to a cosmological
scenario in accordance with well based features of modern
cosmology, as an initial phase dominated by radiation, followed by
a dust epoch long enough to allow structure formation, and by an
accelerated expansion at late times. In addition, by using as
input the present ratio between the vacuum and matter energy
densities, we obtain a universe age in accordance with the
observed limits, and a deceleration parameter that does not differ
from the case of a constant cosmological term.

We have also verified that the matter production characteristic of
the model does not affect the primordial nucleosynthesis, and that
it is not directly observable nowadays. Nevertheless, one could
ask about indirect evidences of this process. As the time scaling
of the energy density during the radiation and dust phases is the
same as in the standard recipe, we do not expect important changes
in what concerns the present temperature and spectrum of the
cosmic microwave background. But it would be an interesting line
of investigation to look for signatures of radiation or matter
production in CMB.

Signatures of matter production may also be found in the relative
abundances of light elements, owing to possible baryon production
originated from the vacuum decay throughout the whole expansion.
On the other hand, the observed abundances may also be used to
establish limits to the decay of vacuum into baryonic matter.
Note, however, that this would not impose limits on the vacuum
decay at all. While we do not have a complete, microscopic
description of the process, we do not know what kind of particles
are produced: baryons, dark matter or any other else. The absence
of a definite quantum field theory for the vacuum decay in curved
backgrounds is also related to another open problem, namely, where
the new matter is generated. Has it some correlation with the
existing matter, or is it produced throughout the entire space?
These questions are common to any cosmological model with matter
production.

As a last point, we have to note that the Sch\"utzhold proposal
does not explain a second important problem related to the
cosmological constant, namely the cosmic coincidence, that is, the
approximate coincidence observed today between the vacuum and
matter densities. This second problem may, in principle, be solved
in other kinds of model with vacuum decay, which consider a
distinct evolution law for $\Lambda$, varying with $H^2$ instead
of a linear relation \cite{others}. These models admit solutions
in which, in the limit of large times, the ratio between the
matter and vacuum energy densities tends to a finite constant. The
problem is that they lead to a different dynamics at early times,
and so are limited by nucleosynthesis and CMB observations.
Another problem is that, in order to get the observed ratio
between the matter and vacuum densities, we obtain a universe age
very high compared to observational bounds \cite{MGX}.

As discussed in other works \cite{MGX,Friedmann}, cosmological
models with decaying vacuum density admit late time solutions
characterized by a constant and correct ratio between vacuum and
matter densities, and in good accordance with the observed
universe age and deceleration parameter. In this case, however, we
need also to suppose a time variation of the gravitational
constant, a hypothesis that depends on observational confirmation,
and that is outside the scope of the present paper.

\section*{Acknowledgments}

The authors are thankful to Ralf Sch\"utzhold for useful
discussions, and to the anonymous referees for constructive
suggestions. This work was partially supported by Fapesb.


\begin{thebibliography}{}
\bibitem{Bertolami2} In this respect, the present model differs from
other approaches (see, for example, reference \cite{Bertolami}),
where the matter energy is conserved separately and the $\Lambda$
variation is followed by a variation of the gravitational
constant.
\end{thebibliography}
\end{document}